\documentclass[prb,aps,twocolumn,showpacs,showkeys,%
 amsmath,amssymb,superscriptaddress,final,reprint,floatfix,longbibliography]{revtex4-1}

\usepackage{lineno}
  \usepackage{mathptmx}
\usepackage{subfigure}
\usepackage{dcolumn}
\usepackage{amsmath,amssymb}
\usepackage{bm}
\usepackage{color}
\usepackage{overpic}
\usepackage{latexsym}
\usepackage{epstopdf}
\usepackage{xcolor}
\usepackage[english]{babel}
\usepackage{latexsym}
\usepackage{stmaryrd}
\usepackage{comment}
\usepackage{tikz}

\usepackage{braket}

\definecolor{mygrey}{gray}{0.35}
\definecolor{myblue}{rgb}{0.2,0.2,0.8}
\definecolor{myzard}{cmyk}{0,0,0.05,0}
\definecolor{mywhite}{rgb}{1,1,1}
\definecolor{myred}{rgb}{1,0.,0.3}

\usepackage[colorlinks=true,citecolor=myblue,linkcolor=myred]{hyperref}
\DeclareMathAlphabet{\mathpzc}{OT1}{pzc}{m}{it}

\def\beq{\begin{equation}}
\def\eeq{\end{equation}}

\def\barray{\begin{eqnarray}}
\def\earray{\end{eqnarray}}

\usepackage{letltxmacro}
\LetLtxMacro{\ORIGselectlanguage}{\selectlanguage}
\makeatletter
\DeclareRobustCommand{\selectlanguage}[1]{%
  \@ifundefined{alias@\string#1}
    {\ORIGselectlanguage{#1}}
    {\begingroup\edef\x{\endgroup
       \noexpand\ORIGselectlanguage{\@nameuse{alias@#1}}}\x}%
}
\newcommand{\definelanguagealias}[2]{%
  \@namedef{alias@#1}{#2}%
}
\makeatother

\DeclareMathOperator{\Tr}{Tr}

\definelanguagealias{en}{english}

\begin{document}

\title{On quenches to the critical point of the three states Potts model - Matrix Product State simulations and CFT}

\author{Niall F. Robertson}
\affiliation{Departament de F\'{\i}sica Qu\`antica i Astrof\'{\i}sica and Institut de Ci\`encies del Cosmos (ICCUB), Universitat de Barcelona,  Mart\'{\i} i Franqu\`es 1, 08028 Barcelona, Catalonia, Spain}
\author{Jacopo Surace}
\affiliation{ICFO-Institut de Ciències Fotòniques, The Barcelona Institute of Science and Technology, 08860 Castelldefels (Barcelona), Spain}

 \author{Luca Tagliacozzo}
\affiliation{Departament de F\'{\i}sica Qu\`antica i Astrof\'{\i}sica and Institut de Ci\`encies del Cosmos (ICCUB), Universitat de Barcelona,  Mart\'{\i} i Franqu\`es 1, 08028 Barcelona, Catalonia, Spain}

\begin{abstract}
Conformal Field Theories (CFTs) have been used extensively to understand the physics of critical lattice models at equilibrium. However, the applicability of CFT calculations to the behaviour of the lattice systems in the out-of-equilibrium setting is not entirely understood. In this work, we compare the CFT results of the evolution of the entanglement spectrum after a quantum quench with numerical calculations of the entanglement spectrum of the three state Potts model using matrix product state simulations. Our results lead us to conjecture that CFT does not describe the entanglement spectrum of the three state Potts model at long times, contrary to what happens in the Ising model. We thus numerically simulate the out-of-equilibrium behaviour of the Potts model according the CFT protocol - i.e. by taking a particular product state and ``cooling'' it, then quenching to the critical point and find that, in this case, the entanglement spectrum is indeed described by the CFT at long times.
\end{abstract}
\maketitle
\section{Introduction}

The simulation of the out-of-equilibrium dynamics of quantum many-body systems is a major challenge with immediate implications for the construction of the next-generation of quantum technologies. In the absence of fully-fledged error correcting quantum computers, tensor-network algorithms and quantum simulators represent some of the best methods currently available to simulate the out-of-equilibrium dynamics of complex quantum systems. Both tools introduce approximations that need to be understood and characterised \cite{hauke_2012}.

Tensor-network algorithms are tailored to slightly entangled states \cite{ran_2020} while typical states that are produced in the out-of-equilibrium dynamics are robustly entangled. These techniques can provide almost exact results only at short times and many fascinating  questions relating to the long-time dynamics of quantum systems out of equilibrium remain open as a result. In particular, it would be interesting to have tools for generic systems that allow the cross-checking of new ideas which aim to predict if a particular system will \textit{thermalise} or not in the long-time limit \cite{pandey_2020}.

Conformal field theories in 2D allow to perform such long-time calculations and thus provide one of the analytical tools to address the out-of-equilibrium dynamics of quantum systems \cite{calabrese_2004,calabrese_2005,calabrese_2006,calabrese_2007,stephan_2011}. Ultimately however, we are interested in the out-of-equilibrium dynamics of lattice models and the CFT only provides an idealised limit of those.

It is well understood how to construct the continuum limit at equilibrium - the low energy part of a system that is at its critical point is described by the underlying CFT \cite{francesco_1997,henkel_1999}. In the out-of-equilibrium setting, the relation between the CFT and lattice models  is less clear. A standard quench protocol indeed not only probes the low energy physics of a system but also the physics in the middle of the spectrum \cite{polkovnikov_2011,haldar_2020}. It is thus not obvious what part of the lattice physics should thus be described by the CFT. 

We can still hope that robust, universal quantities, on the lattice, could be described (in a certain limit to be specified below) by the CFT. The hope is  substantiated by the previous results  obtained  simulating the long-time behaviour of quantum systems out-of-equilibrium \cite{giulio_2019,surace_2019,surace_2020} and comparing them with the CFT predictions. In previous works \cite{torlai_2014,giulio_2019,surace_2020} we have identified the low energy part of the entanglement Hamiltonian as the relevant quantity to focus on in this context. We have shown how the CFT predictions are recovered by focusing on the low-energy part of the entanglement Hamiltonian. In fact, it has been shown \cite{torlai_2014} that when quenching to a critical point (and across it), the entanglement Hamiltonian (to be defined below) becomes gapless and its low energy part is appropriately described by a CFT.

Being able to identify universal effects is very important, since it would allow the study of deviations from the universal behaviour by means of a scaling theory.  For example,  at equilibrium, one can build various scaling forms that allow to characterise criticality with approximate tensor networks techniques. The idea there is that  the approximations induced by the numerical techniques can be mapped to relevant deformations of the underlying CFT \cite{nishino_1996,tagliacozzo_2008,pollmann_2009,pirvu_2012,stojevic_2015,corboz_2018,rader_2018,rams_2018,czarnik_2019,vanhecke_2019}. 

Here we focus on the three-state Potts model on the lattice. Based on the CFT calculations \cite{cardy_2016} and on our results for the Ising model, one might expect that by quenching an initial state of the lattice model  to the critical point, the low-energy part of the entanglement Hamiltonian of a region should encode the spectrum of the three-state Potts CFT on an annulus. 

This is not what we observe. In our numerical results we find strong deviations from the desired spectrum. We still observe that during the out-of-equilibrium dynamics after a quench to the critical point, the entanglement Hamiltonian becomes gapless.  Its spectrum however significantly deviates from the one expected for the three states Potts CFT with free boundary conditions (and by a matter of fact with any kind of conformally invariant boundary conditions). We try to understand if the deviations we observe are due to the presence of strong corrections to scaling. 
At equilibrium, indeed there are strong corrections to the CFT predictions for the entanglement spectrum caused by the finite size of both the region and the system. As a result, and in order to recover the CFT predictions, one must perform appropriate finite size extrapolations as already pointed out by \cite{lauchli_2013}. For the spectrum at equilibrium we show that, by using insights from CFT, we can actually perform such extrapolations in practice and recover precisely the expected analytical results.

We also show how, for the Ising model, similar extrapolations allow to extract the out-of-equilibrium CFT predictions of the entanglement spectrum from the data of the lattice model using only very small systems and short times. We finally report the failure of such a strategy in the case of the out-of-equilibrium dynamics of the three states Potts model. 
None of the extrapolations we attempted reproduce the expected CFT spectrum. This suggests that the effects observed on the lattice in the out-of-equilibrium dynamics of the Potts model are not solely a result of irrelevant perturbations that can be accounted for by simple extrapolations.

We conclude by showing that the CFT results for the Potts dynamics can be obtained from the lattice by simulating a different out-of-equilibrium protocol consisting of initially ``cooling'' a product state (the lattice version of a conformally invariant boundary state) and then quenching the system using its critical Hamiltonian. This protocol is a direct simulation of the prescription used to perform the CFT calculations as we review in the following. Our work thus sheds some light on the regime of applicability and limitations of the CFT analysis to the out-of-equilibrium dynamics of lattice models after a quench.

\section{Definitions and review of the CFT approach to quenches}
\label{sec:CFT}

\begin{figure}
	\centering
\begin{tikzpicture}[scale=1.4]
\draw[black,line width = 1pt](0.5,0.5)--(5.5,0.5);
\draw[black,line width = 1pt](0.5,2.5)--(5.5,2.5);
\draw[black,line width = 1pt](0.5,0.5)--(0.5,2.5);

\draw[black,line width = 1pt,dashed](0.5,1.65)--(2.5,1.65);
\draw[black,line width = 1pt,dashed](0.5,1.55)--(2.5,1.55);


\begin{scope}
	\clip (2.5, 1.3) rectangle (2.8, 1.7);
\draw[black,line width = 1pt](2.58,1.6) circle(0.1);
\end{scope}

\node at (2.9,1.6) {$\ket{a}$};
\node at (0.2,1.5) {$\ket{b}$};
\node at (2.8,0.3) {$\ket{b}$};
\node at (2.8,2.7) {$\ket{b}$};

\end{tikzpicture}

\caption{The semi-infinite strip of width $\frac{\beta}{2}$ with an entangling cut at $\ket{a}$. The boundaries of the strip are mapped to the outer rim of the annulus under the conformal transformation, and the small disk at the entangling cut is mapped to the inner rim. The cut connecting $\ket{a}$ to $\ket{b}$ should be vertically offset by an amount proportional to $\tau$ from the  middle of the strip.}\label{cftpic}
\end{figure}
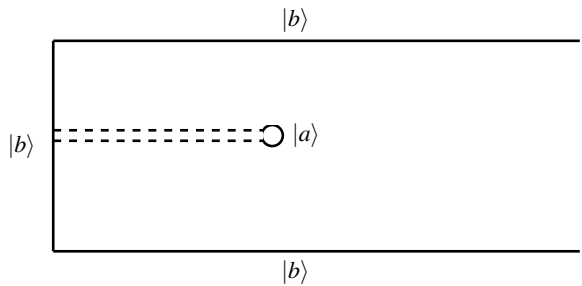

The entanglement Hamiltonian is defined as follows: if we make a bipartition of our quantum system into subsystems $A$ and $B$, the entanglement Hamiltonian of the subsystem $A$ is defined via the reduced density matrix $\rho_A$ as
\beq\label{entanghamdef}
\mathcal{H}_A = -\frac{1}{2\pi}\ln \rho_A,
\eeq
where
\beq\label{rhodef}
\rho_A = \Tr_B \ket{\psi} \bra{\psi},
\eeq
where $\Tr_B$ is the partial trace over $B$. If the system is at zero temperature then $\ket{\psi}$ is taken to be the ground state of the system. We refer to the spectrum of $\mathcal{H}_A$ as the entanglement spectrum. A paradigmatic example of an out-of-equilibrium protocol is a quantum quench, whereby the system is initially prepared in a state $\ket{\psi_0}$ - taken to be the ground state of some Hamiltonian - and is subsequently time evolved with a different Hamiltonian. In our case, the Hamiltonian governing the time-evolution will be the Hamiltonian at the critical point, or in the field theory approach, that of the Conformal Field Theory $H_{\text{CFT}}$. We will be particularly interested in the ratios
\beq\label{ratiosdef}
r_j = \frac{\epsilon_j - \epsilon_0}{\epsilon_1 - \epsilon_0},
\eeq
where $\epsilon_j$ are the eigenvalues of the entanglement Hamiltonian. The CFT calculations tell us that when one considers the entanglement Hamiltonian of a finite block\cite{wen_2018} or of a semi-infinite line\cite{cardy_2016}, the spectrum should be equal to that of the CFT Hamiltonian up to multiplicative and additive renormalisation constants. This is related to the fact that  conformal transformations exist that map the time evolution operator of CFTs in certain geometries to the entanglement Hamiltonian of a subregion of the system \cite{wen_2016}. The ratios $r_j$ should thus coincide with the ratios $\frac{h_j-h_0}{h_1-h_0}$ where $h_j$ are the conformal dimensions of the corresponding boundary CFT. The strategies for both CFT calculations, i.e. for the calculation of the entanglement Hamiltonian of a finite block and of a semi-infinite line respectively, are broadly the same; one starts with the assumption that the initial state is given by 
$\ket{\psi_0} = e^{-\frac{\beta}{4} H_{\text{CFT}}}\ket{b}$
(the so-called ``Cardy-Calabrese'' state) where $\ket{b}$ is a conformally invariant boundary state with zero entanglement. Evolving with a small amount of imaginary time $\beta$ introduces a small amount of entanglement to the system. One can interpret $\beta$ as the correlation length that is introduced. One can then write the density matrix $\rho$ in Euclidean time as
\beq\label{rhotau}
\rho(\tau) \propto e^{-H\tau}e^{-\frac{\beta}{4} H_{\text{CFT}}}\ket{b}\bra{b} e^{-\frac{\beta}{4} H_{\text{CFT}}}e^{H\tau},
\eeq
where one takes $\tau\rightarrow it$ at the end of the calculation. Equation (\ref{rhotau}) is then interpreted as the path integral of strip of width $\frac{\beta}{2}$ with a branch cut along a straight line connecting the boundary and the entanglement cut (that is, between $\ket{b}$ and  $\ket{a}$ in Figure \ref{cftpic}). Then the strip (with a small disc removed at the entangling point to regularise the theory \cite{wen_2018, cardy_2016}) is mapped to the annulus by a particular conformal mapping $w = f(z)$ such that the branch cut now connects the two boundaries on the annulus. The entanglement Hamiltonian on the strip is mapped to the generator of translations around the annulus. Its spectrum is thus given (up to an overall shift and rescaling) by that of a boundary CFT on the annulus: $\frac{\pi(h_j - \frac{c}{24})}{W}$ where $h_j$ are the conformal dimensions of the boundary CFT, $c$ is the central charge and $W$ is the width of the annulus obtained from the conformal transformation. There are two conformal boundary conditions to consider - on the inner and outer rim of the annulus respectively. The inner rim corresponds to the point $a$ in the strip at which the entangling cut was made and usually corresponds to free boundary conditions. The outer rim could in principle have two separate conformal boundary conditions on two different parts, corresponding to the vertical and horizontal solid lines in Figure \ref{cftpic}. Here we will only consider the case where these boundary conditions are all identical. \\

The CFT calculations that have just been outlined have recently been shown \cite{surace_2020} to predict the out-of-equilibrium dynamics of the entanglement spectrum of the quantum Ising chain at long times. In particular, it was verified that for the transverse field Ising model, the low energy part of the entanglement Hamiltonian encodes the universal spectrum of the Ising CFT on an annulus. Similar results have been also presented in the context of free bosons and fermions \cite{giulio_2019}.\\

Note also that the CFT calculations predict two different regimes in the time evolution after a quench to the critical point. A first regime where the entanglement entropy grows linearly and a second regime where it saturates to a value proportional to the volume of the region \cite{calabrese_2005,calabrese_2006,calabrese_2007}.
Furthermore, since we deal with finite systems, there are also other regimes that occur after saturation, where the entropy starts to decrease and then increases again thus giving rise to quantum recurrences. We will not deal with them here and we will mainly focus on the first two regimes, the initial regime of entropy growth followed by saturation. These are the only two regimes that would survive in the thermodynamic limit. 

In the Ising model, the numerical results for both regimes agree with the analytical predictions of the  CFT analysis, and one can justify the agreement by considering different limits of the CFT analysis. For example, in the regime of linear growth one can just consider short enough times compared to the times needed for the information to reach the boundaries of the system. This regime should be described by the CFT analysis that considers half an infinite chain embedded in an infinite system. Due to the locality of interactions and the short times there should be little or no effects induced by the finite size of the region and system \cite{surace_2020}.

In the regime of entanglement saturation, we can invoke (general) thermalization and the corresponding finite correlation length. The finite correlation length once more allows the description of a finite region embedded in a finite system by the CFT calculation for an infinite region embedded in an infinite thermal system with a finite correlation length.
As long as the size of the region and system exceed the thermal correlation length, we expect that the CFT description should be exponentially close to that of the finite system \cite{surace_2020}.

The last paragraph is mostly justified by the numerical observation made in  \cite{surace_2020} that show that, in the continuum limit, the low energy part of the entanglement Hamiltonian of a region in a Gibbs state is indistinguishable from the low energy part of the entanglement Hamiltonian of a region in a generalized Gibbs state.


\section{The three states Potts model and its underlying CFT}\label{setup}
We will consider the three state Potts model with open boundary conditions. The Hamiltonian is given by:
\beq\label{hpotts}
H = -J\sum\limits_{i=1}^{N-1}\left( \sigma_i\sigma_{i+1}^{\dagger} + \sigma_i^{\dagger}\sigma_{i+1} \right) - f\sum\limits_{i=1}^{N}(\tau_i + \tau_i^{\dagger}).
\eeq
The operators are $\sigma=\sum_{s=0\cdots 2} \omega^s \ket{s}\bra{s}$ with $\omega = e^{i2\pi/3}$, while $\tau=\sum_{s=0\cdots 2} \ket{s}\bra{s+1}$ where the addition is considered modulo 3. The system is disordered for $f > J$, ordered for $f < J$ and is critical for $J=f$ where the symmetry is spontaneously broken \cite{fendley_2012,mong_2014}. The CFT describing this critical point is the minimal model with central charge $c=4/5$. Let's first briefly recall some of the other key results of the boundary CFT that describes the model. The full generating function of levels for the model with free boundary conditions on both sides is given by \cite{affleck_1998}
\beq\label{Zfreefree}
Z_{\text{free, free}} = \chi_{I} + \chi_{\psi} + \chi_{\psi^{\dagger}},
\eeq
where $\chi_{I}$, $\chi_{\psi}$ and $\chi_{\psi^{\dagger}}$ can be written in terms of generators  $\chi_{r,s}$ of the Kac modules:
\beq
\chi_{\psi} = \chi_{\psi^{\dagger}} = \chi_{43},
\eeq
and
\beq
\chi_{I} = \chi_{11} + \chi_{41}.
\eeq
The corresponding conformal dimensions $h_{r,s}$ are given by $h_{11} = 0$, $h_{41} = 3$ and $h_{43} = \frac{2}{3}$. 
and the explicit form of the generating functions are:
\beq\label{genfuncs}
\begin{aligned}
\chi_{11} &=  q^{-c/24} \left(1 + q^2 + q^3 + 2q^4 + 2q^5 + 4q^6 + ...\right),\\
\chi_{41} &= q^{3-c/24}\left( 1 + q + 2q^2 + 3q^3 + 4q^4 + 5q^5 + ...\right),\\
\chi_{43} &= q^{2/3-c/24}\left( 1 + q + 2q^2 + 2q^3 + 4q^4 + 5q^5 + ...\right).
\end{aligned}
\eeq
The generating functions in (\ref{genfuncs}) give us the degeneracies of the descendants of each primary state. As mentioned in the introduction, one only expects the entanglement spectrum to be equal to the CFT spectrum up to an overall shift and rescaling. To deal with this issue, we compare the ratio of the gaps of the entanglement spectrum with those of the CFT, thus eliminating both the additive and multiplicative constants.

More precisely, we expect the ratios of the gaps of the entanglement spectrum, defined in equation (\ref{ratiosdef}), to converge to the corresponding ratios of the CFT, given by $\frac{h_i-h_0}{h_1-h_0}$. From equation (\ref{Zfreefree}) we thus expect these ratios to be: $\{1, \frac{5}{2}, \frac{5}{2}, 3, 4,4,4,4, \frac{9}{2}, \frac{9}{2},...  \}$ where the degeneracies come from the coefficients of the terms in (\ref{genfuncs}).  

\begin{figure}[!htb]
\includegraphics[scale=0.5]{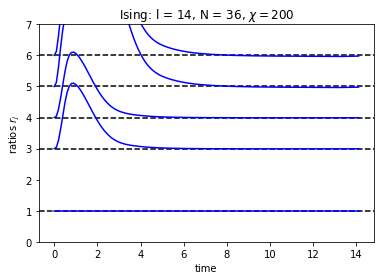}
  \caption{The ratio of the gaps in the entanglement spectrum as a function of time for a quench from $g = 1.5$ to the critical point $g = 0.5$ in the Ising model. The length of the chain is taken to be $N=36$ and the length of the block is $l=14$. Despite the very modest size of the system and bipartition we see that the CFT predictions (dotted lines) are accurately reproduced by the ratio of the numerically studied entanglement spectrum ratios. During the time interval considered in the figure the entropy is always increasing linearly.}\label{quenchIsing} 
\end{figure}

In addition to the three-states Potts model, we will also consider here the Ising model in a transverse field by way of comparison. The Ising Hamiltonian is given by
\beq\label{hising}
H = -J\sum\limits_{i=1}^{N-1} \sigma_i^z \sigma_{i+1}^z -g\sum\limits_{i=1}^{N} \sigma_i^x.
\eeq
The model is critical for $J=g$ and in what follows we take $J=\frac{1}{2}$, so that the critical point is given by $g = \frac{1}{2}$. In Figure \ref{quenchIsing} we plot the ratio of the gaps of the entanglement spectrum vs time after a quench from the disordered regime ($g > J$) to the critical point. 

\section{Numerical results}
\subsection{Results at equilibrium}\label{equilresults}
\begin{figure}[!htb]
 	\includegraphics[scale=0.5]{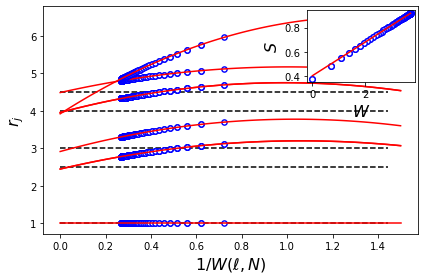}
  \caption{The ratio of the gaps in the entanglement spectrum of the three state Potts model in the ground state at the critical point. The numerical ratios $r_j$ are the blue dots defined in (\ref{ratiosdef}). The red lines are best-fit to the data including corrections up to $1/W^2$. We see that the results of the fit  converge to the expected CFT values (represented by dashed horizontal lines) as $\frac{1}{W} \rightarrow 0$. Here our numerical results are obtained by considering all possible bipartitions of a $N=128$ spin chain in two complementary blocks of $l$ and $N-l$ consecutive spins. The bond dimension of the ground state corresponding MPS is $D=585$ in order to ensure almost exact results. The inset shows that in the same range of values of $W$, the entropy grows linearly with respect to $W$ as expected.}\label{PottsEquilib} 
 \end{figure}

Before discussing the out-of-equilibrium dynamics of the entanglement spectrum it is important to review some of the well known results for the equilibrium case. The CFT analysis is very similar to that of the out-of-equilibrium case discussed previously and involves conformally mapping the reduced density matrix to a path integral on a strip. Numerical results from Lauchli \cite{lauchli_2013} showed that the entanglement spectrum of both the Ising model and the three state Potts model at equilibrium is given by the corresponding CFT spectrum. To see this one needs to undertake a finite size scaling analysis and extrapolate to $W \rightarrow \infty$ where $W$ is the width of the annulus to which the system is conformally mapped. One has that $W = \log\left( \frac{2N}{\pi} sin(\frac{\pi l}{N})\right)$ \cite{calabrese_2004, roy_2020} where we recall that $N$ is the length of the system and $l$ is the length of the block under consideration. In Figure \ref{PottsEquilib} we show the finite size scaling results for the entanglement spectrum of the Potts model at equilibrium. The system we consider is already relatively large, being made up of $N=128$ spins. The entanglement spectra are obtained for all possible bipartition of the system into two blocks of consecutive spins, one of size $l$ and the other of size $N-l$ and the corresponding ratios $r_j$ are plotted as empty circles as a function of $W(l,N)$. We see that, although there are large corrections to the leading scaling behavior, when plotted as a function of $W$ we can extrapolate the numerical results to the expected CFT spectrum. In particular the CFT spectrum  is recovered in the limit $\frac{1}{W} \rightarrow 0$, including all of the correct degeneracies predicted by (\ref{genfuncs}) by including effects up to $1/W^2$. The corresponding fits are presented as solid red lines that (almost) intersect the CFT predictions at $1/W=0$. Notice that such precise extrapolation is impossible  without knowing the explicit form of $W$  (see e.g. the lines traced in \cite{lauchli_2013} that have to be considered as guide to the eyes rather than exact extrapolations). 

\subsection{Results out of equilibrium}
\begin{figure}[!htb]
	\centering
	\includegraphics[scale=0.5]{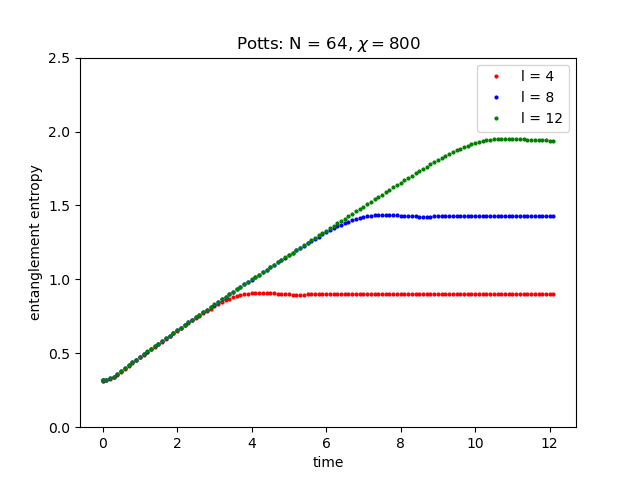}
	\includegraphics[scale=0.55]{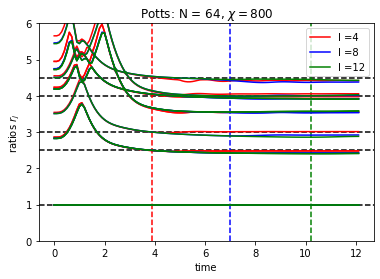}
 \caption{\textit{Upper panel}: Entropies vs time after a quantum quench from $f = 0.6$ to the critical point $f = 0.5$. Blocks of length $l=4, 8, 12$ are considered in a chain of length $N = 64$. We see a regime where the entropy grows linearly followed by a regime where the entropy is saturated. \textit{Lower panel}: the ratios of the gaps in the entanglement spectrum of the Potts model as a function of time for a quench from $f=0.6$ to the critical point $f=0.5$. We see that the numerics deviate from the CFT ratios (horizontal dotted lines). In particular, there is a ratio at roughly $r = 3.5$ which cannot be predicted by the CFT calculation. The vertical dotted lines mark the points at which the entanglement entropy of the respective blocks saturates.}\label{quenchpotts1}

\end{figure}

We would like to apply a similar analysis to the out-of-equilibrium case. We start by focusing on the short time regime of linear growth of entanglement. The regime is easily identified in the upper panel of Fig. \ref{quenchpotts1} where we  plot the entanglement entropy of bipartitions of increasing size as a function of time. The regime we consider is that with times shorter than the entropy saturation time. Such saturation times are represented by vertical lines in the lower panel of Fig. \ref{quenchpotts1}, with the color corresponding to the desired subsystem sizes. In the lower panel  we show the typical behaviour of the ratios $r_j$ of the entanglement spectrum of a partition of $l=16$ in a system of $N=64$ described with an MPS of bond dimension up to $D=800$ \footnote{We have checked that the results are the same also for larger bond dimension of the order of $D=2000$.}. The horizontal lines in the figure highlight the predictions from the CFT while the solid lines in different colours show the time evolution of the various ratios $r_j$ for different block size $l$.  The plot presents a typical quench from $f=0.6$ in the disorder phase to the critical point, but all the other quenches to the critical point we have analysed  show a similar behaviour. As discussed previously, the Hamiltonian in (\ref{hpotts}) is critical for $J=f$. In what follows, we always take $J=\frac{1}{2}$ so that the critical point is given by $f=\frac{1}{2}$. It is clear that the obtained data do not correspond to the CFT calculation, as seen by the fact that the solid curves clearly deviate away from the CFT predictions. Furthermore, with this particular choice of the initial state, the numerical data suggest the appearance of an extra spurious ratio at around $3.5$, but the further we move away the initial state from the critical point the larger the deviations between the numerical spectrum and the CFT predictions become and this spurious level moves to different values.

Also notice that the trend of the lines when we increase the  block sizes does not follow what we would expect. The spectrum for larger blocks is further away from the CFT predictions than the one of smaller blocks. Strictly speaking increasing the size of the block does not correspond to taking the continuum limit, that in this first regime should be obtained by taking $t \to \infty$, $\beta \to \infty$ by keeping the ratio $t / \beta$ constant. However, at fixed $\beta$,  in the short time limit, larger blocks allow to get to larger times before the entropy saturates and thus allow to explore larger values of the ratio. 

Comparing this situation with the one at equilibrium one might argue that this apparent discrepancy can be cured by performing an appropriate finite $W$ extrapolation. The main obstacle to performing such an extrapolation is the lack of knowledge of the exact formula for $W$ in the context of the out-of-equilibrium dynamics that we are considering here. An expression for $W$ can only be obtained when the interval is a semi-infinite part of an infinite chain where $W(t, \beta)=\log\left( \cosh (\frac{2 \pi t}{\beta})\right)$ \cite{cardy_2016}. As already mentioned this expression can only be used in the regime of linear entanglement growth where we expect that, due to the finite speed of the propagation of the information, there is no explicit finite size dependence of the physics. An expression for $W$ for a finite block at the end of a semi-infinite chain is also available in \cite{wen_2018}. In the limit $t, l >> \beta$, both expressions can be simplified giving rise to $W \propto t$. As a result, at fixed $\beta$, that is when considering quenches from a fixed initial state, the large $W$ limit coincides with the large $t$ limit.

Unfortunately we cannot consider $t$ arbitrarily large  since as we mentioned, the computational complexity of the MPS simulations increases exponentially with time, and thus with $W$ \footnote{Notice that this is the case also at equilibrium, the computational complexity for the ground state of a critical system scales also linearly with $W$ that in this case is $W \propto \log(L)$ rather than $W\propto t$. Interestingly at equilibrium values of $W\simeq 10$ are enough to extract the expected results, while as we show here, similar values of $W$ are insufficient to extract the CFT predictions here}. Also, in order to remain in the regime of linear growth of the entropy, the finite size of the blocks we consider provides an upper bound to the largest $t$ we can address, as shown explicitly in the upper panel of Fig. \ref{quenchpotts1} \cite{calabrese_2005,surace_2020}. 
As a result, due to our computational limitations, we can only study accurately relatively short times and we would not gain anything by simulating larger than $N=64$ spins. With $N=64$ we can safely accommodate blocks of size $l\le 12$  which stay in the regime of linear growth for all the times we are able to simulate accurately (typically of the order $t/J\simeq 10$) as shown in the top panel of Fig. \ref{quenchpotts1}.


Since the times we can consider are still large enough to access the regime,  $\beta < t < l$ we can try to use the results we obtain in order to extrapolate them to the infinite-time limit. Before performing these extrapolations on the the Potts model's results we benchmark our extrapolation strategy for the  Ising model where, by using the exact mapping to free fermions, we have shown that, for sufficiently large times and systems, the CFT emerges without any need of extrapolation \cite{surace_2020}. The Hamiltonian is defined in Eq. \eqref{hising}.

\begin{figure}[!htb]
	\includegraphics[scale=0.5]{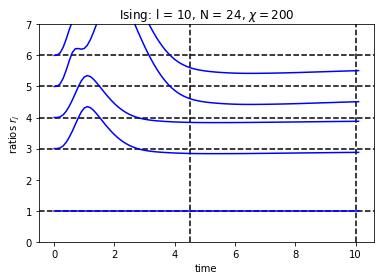} 
	\includegraphics[scale=0.5]{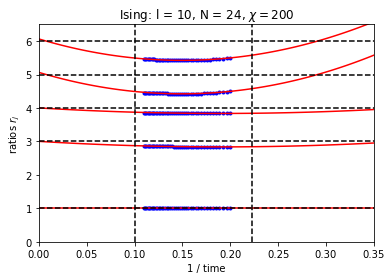}
 \caption{\textit{Upper panel}: The ratio of the gaps in the entanglement spectrum as a function of time for a quench from $g = 0.8$ to the critical point $g = 0.5$ in the Ising model. One observes that for the time scales considered here the ratios do not reach their CFT values given by the horizontal dotted lines. However, these ratios are expected to reach the horizontal dotted lines after long enough times. In the time interval considered here we have checked that the entropy is still in the regime of linear growth. The vertical dotted lines in both figures represent the time window used  to extrapolate to long times (see lower panel). \textit{Lower panel}: The finite-time scaling of the ratios of the gaps in the Ising model. The ratios $r_j$ are plotted vs $\frac{1}{\text{t}}$ because, as discussed in the main text, $W \propto t$. After extrapolation, the ratios $r_j$ converge to their CFT values (the horizontal dotted lines). The extrapolation is stable for any subset of data points inside the time window. }\label{fssIsing} 
\end{figure}

The numerical results for the various $r_j$ for a quench from $g=1.5$ to the critical point are presented as solid lines while the analytical predictions are presented as dotted horizontal lines.  Notice that the system is smaller than the system we are studying in the Potts case, namely made by $N=36$ spins.  Nevertheless we see that the ratios of the gaps of the entanglement spectrum rapidly converge to their CFT values (the dotted lines), as has been previously observed \cite{surace_2020}. The fact that corrections to scaling are much smaller in the Ising case w.r.t to the Potts case should not come as a surprise and was already observed in \cite{lauchli_2013} at equilibrium. 
In order to further amplify the finite-time regime in the Ising case we need to reduce the system size down to $N=24$ spins.  In Figure \ref{fssIsing} we plot the ratio of the entanglement gaps $r_j$ vs time after a quench in the Ising model for short times such that the gaps have not yet reached their CFT values. The numerical results are encoded by red dots, while the CFT predictions are dashed horizontal lines. 

Using $W \propto \frac{1}{t}$, we plot the ratio of the gaps vs $\frac{1}{t}$ and extrapolate to $\frac{1}{t} \rightarrow 0$ using a quadratic polynomial in $1/t$.  The corresponding best fit are shown by solid lines in colors, and  (almost) intersect the  CFT predictions at $1/t=0$. Notice that there are several potential causes of error involved in this analysis that one must be careful to take into account. The most difficult to control is the one associated with the choice of the time window. We need to consider large enough times such that the CFT starts to emerge at low-entanglement energies but short enough to stay away from the regime of entanglement saturation.  Figure \ref{fssIsing} shows the result of the best window that allows to extract the CFT predictions in the limit $\frac{1}{t} \rightarrow 0$.
Now we can attempt the same extrapolation for the quench in the Potts model presented in Fig. \ref{quenchpotts1}. 

\begin{figure}[!htb]
	\includegraphics[scale=0.5]{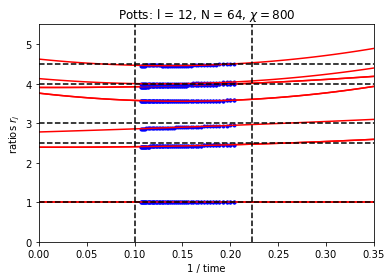}
 \caption{The finite size scaling of the ratios of the gaps $r_j$ in the Potts model. Unlike the Ising model in Figure \ref{fssIsing}, the ratios $r_j$ do not appear to converge to the CFT values even after extrapolation. As in Figure \ref{fssIsing}, the vertical dotted lines represent the extremes of the time interval in which we use the data points in the extrapolation.\label{fssPotts}} 
\end{figure}

In Figure \ref{fssPotts} we repeat the same analysis we have performed for the Ising model for Potts. There, despite considering larger systems sizes and blocks than those of Ising,  the same finite-time scaling analysis fails to unveil the spectrum of the three state Potts CFT. The numerical results are presented once more by red lines, the CFT predictions by horizontal dashed lines and the extrapolations, performed by using second order polynomial in $1/t$, are presented by solid lines of different colours.  One would thus be tempted to assert that the results of the CFT cannot be obtained by considering corrections to the scaling of the data obtained from the numerical quenches in the Potts model.  A word of caution is however necessary, it is important to notice, that all finite size extrapolations we have done so far at equilibrium rely on the exact knowledge of $W$.

We thus try to use another bit of input from the CFT and take seriously the definition of $W$ reported above obtained for  infinite systems. That definition requires the determination of $\beta$, that encodes the dependence of the quench results from the initial state.  One can extract $\beta$ from the simplest CFT predictions about the linear growth of the entanglement entropy with $W$, $S=c/12\  W (t,\beta)$. Once we assume the expected $c=4/5$, $\beta$ is determined by adjusting it so that a linear fit to the entropy provides the correct prefactor $c/12$  \cite{coser_2014a}. Knowing $\beta$ we can thus compute $W(t, \beta)$. 

\begin{figure} 
 \includegraphics[width=\columnwidth]{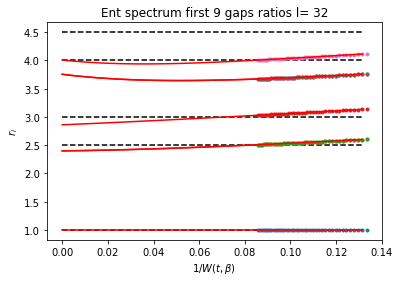}
 \caption{By using $\beta$ extracted from the scaling of the entanglement entropy we can plot the ratios $r_j$ as a function of $W$ calculated within the CFT in the thermodynamic limit. By accurately selecting the correct time window to ensure that we are in the intermediate time regime, we can try to perform a similar extrapolation to the one we have performed for the equilibrium case in Fig. \ref{equilresults}. Unfortunately the extrapolation does not allow to recover the expected CFT predictions. Here the quench is performed from $f=0.7$ to $f=0.5$.\label{fig:potts_vs_w} }
\end{figure}

This allows us to study  the entanglement gaps ratios $r_j$ as a function of $1/W$, in the limit in which $t\ll l$, that can be ensured by appropriately checking that the growth of the entropy is still proportional to $W$. The results of this refined extrapolation are presented in Fig. \ref{fig:potts_vs_w} where we appreciate that now two of the  extrapolations fall relatively close to the expected $r_2=2.5$ and $r_3=3.0$ but we still observe a spurious ratio at around $r_4=3.5$ that is not part of the known spectra for the three-states Potts CFT.

%
%


The failure of our extrapolations lead us to conjecture that the entanglement spectrum observed after a quench in the three states Potts model in the regime of linear growth,  will never reach the CFT values \footnote{We have carefully checked that all possible systematic errors we could introduce by encoding the state with a MPS (finite bond dimensions of the MPS, finite Trotter steps) are not at the origin of the mismatch.} Similar analysis repeated in the regime of entanglement saturation provide similar results.

As mentioned in Sect. \ref{sec:CFT}, the difference between our numerical results and the CFT calculations does not imply that those calculations are wrong. In fact, as we will show in \ref{cftprotocol}, if we \textit{explicitly} simulate the exact CFT protocol on the lattice by evolving a boundary state with a small amount of imaginary time on the lattice, then the ratios of the gaps of the entanglement spectrum are in excellent agreement with the CFT results.
\subsection{Numerical results obtained by simulating the CFT protocol}\label{cftprotocol}
\begin{figure}[ht]
	\includegraphics[scale = 0.5]{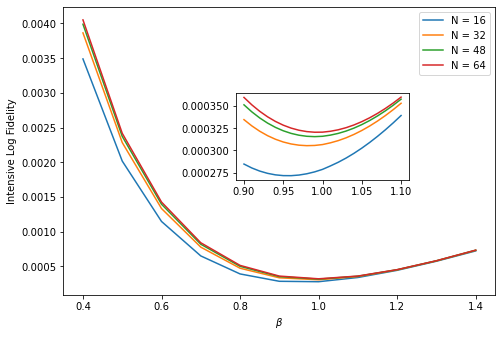}
 \caption{Plot of the intensive logarithmic fidelity of the initial state with the ground state as a function of $\beta$, with $f = 0.6$.}\label{overlapvsbeta}
\end{figure}
\begin{figure}[ht]
	\includegraphics[scale = 0.5]{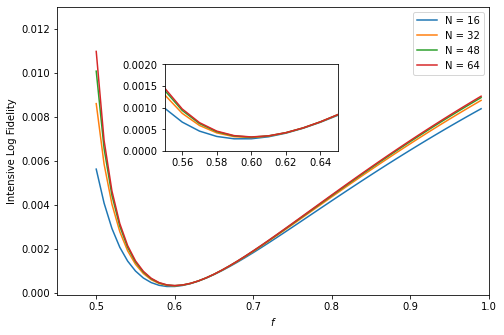}
 \caption{Plot of the intensive logarithmic fidelity of the initial state with the ground state as a function of $f$, with $\beta = 1.0$.}\label{overlapvsf}
\end{figure}
As discussed in the introduction, the initial state in the CFT calculation has the form $\ket{\psi_0} = e^{-\frac{\beta}{4} H_{\text{CFT}}}\ket{b}$ where $\ket{b}$ is a conformally invariant boundary state with zero entanglement and $\beta$ is the correlation length that is introduced as a result of the imaginary time evolution. Consider the state $\ket{b}$ corresponding to free boundary conditions. If we label the three Potts `spins' by $2$, $1$ and $0$ then $\ket{b}$ can be written on the lattice explicitly as: 
\beq\label{freebc}
\ket{b} = [\frac{1}{\sqrt{3}}(\ket{2} + \ket{1} + \ket{0})]^{\otimes N},
\eeq
where $N$ is the length of the chain. We then explicitly construct the initial state $\ket{\psi_0}$ by evolving $\ket{b}$ with imaginary time to obtain $e^{-\frac{\beta}{4} H_{\text{CFT}}}\ket{b}$. In Figures \ref{overlapvsbeta} and \ref{overlapvsf} we plot the logarithmic intensive fidelity of this state with the ground state of the Hamiltonian in (\ref{hpotts}) as a function of $\beta$ and $f$ respectively. More precisely, we plot the quantity $-\frac{1}{N}{\log \bra{\psi_{gr}(f)} e^{-\frac{\beta}{4} H_{\text{CFT}}}  \ket{b}}$ vs $\beta$ and $f$ respectively. \\

\begin{figure}
	\includegraphics[scale = 0.5]{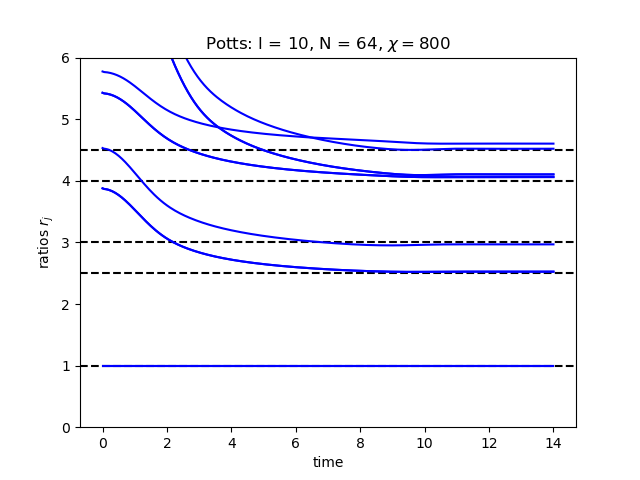}
	\includegraphics[scale = 0.5]{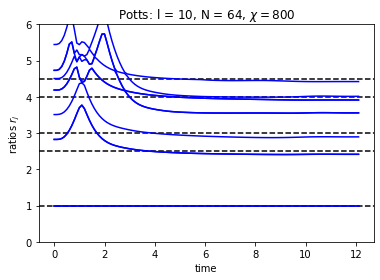}
 \caption{\textit{Upper Panel}: The ratio of the gaps in the entanglement spectrum as a function of time when the initial state is $\ket{\psi_0} = \exp(-H \beta)\ket{b}$ with $\beta = 1.0$, $N=64$, $l=10$, and the state is subsequently time evolved with the critical Potts Hamiltonian. We see close agreement between the numerical results and the CFT results. \textit{Lower Panel}: The ratio of the gaps in the entanglement spectrum as a function of time after a quench from $f = 0.6$ to the critical point $f = 0.5$. As we saw in Figure \ref{quenchpotts1}, the numerical results do not agree with the CFT calculation after the quantum quench.}\label{imagtimebeta1_pt_0}
\end{figure}

\noindent One observes that for each value of $\beta$($f$), there is a corresponding value of $f$ ($\beta$) such that the intensive logarithmic fidelity is very close to $0$. This is in agreement with the expectation that $\exp(-H_{\text{CFT}} \beta)\ket{b}$ should approximate the gapped Hamiltonian. The intensive logarithmic fidelity stays finite in the limit $N \rightarrow \infty$ making it possible that the entanglement spectrum of the block of length $l$ results different in the two cases $\ket{\psi_0} = \ket{\psi_{gr}}$ and $\ket{\psi_0} = e^{-\frac{\beta}{4} H_{\text{CFT}}}\ket{b}$. Observe from Figure \ref{overlapvsbeta} that when we fix $f=0.6$, the point of maximum fidelity (i.e. the minimum of the curve) occurs at a value of $\beta \approx 1.0$. Note from the inset that this value of $\beta$ increases with system size $N$. In Figure \ref{imagtimebeta1_pt_0}, we thus compare the dynamical behaviour of the ratios of the gaps of the entanglement spectrum for the two initial states. Despite the high fidelity of the initial states, it is only for the case $\ket{\psi_0} = e^{-\frac{\beta}{4} H_{\text{CFT}}}\ket{b}$ that one finds good agreement with the CFT predicition. In this case all of the ratios converge to the CFT values marked by the dotted lines.
\section{Discussion}
In this work we have observed the following: i) the CFT analysis in the out-of-equilibrium setting does not appear to predict the behaviour of the entanglement spectrum after a quench to the critical point. ii) the CFT analysis does indeed describe the behaviour of the entanglement spectrum on the lattice when we explicitly simulate the CFT protocol. 

The mildest conclusion is that, for the case of the quantum quench, there are large deviations from the scaling behaviour that one obtains from the CFT calculation. Our multiple attempts to extrapolate the numerical results by including the leading corrections to scaling terms up to $1/W^2$ (sufficient both at equilibrium and for the Ising model) failed. This seems to suggest that the discrepancy is not due to corrections to scaling. We note however that this statement is just a conjecture for now.

Here we elaborate on some of the possible reasons for the observed difference between the two types of quenches that we studied here. We leave a more in depth analysis of each of them for future work.

To apply the CFT calculation to the quantum quench on the lattice, one must assume that the simple state obtained by evolving the conformal boundary state in Euclidean time (i.e. the Cardy-Calabrese state) is enough to describe the results after the quench. This assumption has been analysed in depth in \cite{cardy_2016a}, since it leads to thermalisation rather than to generalised thermalisation. In \cite{cardy_2016a} several proposals were put forward as generalisations of that simple initial state. One could try to design the appropriate initial state that reproduces the results that we observe numerically.

Previously, we have shown that at low-entanglement energies, thermalisation and generalised thermalisation actually coincide for the Ising model \cite{surace_2020}. This was visible at very small system sizes as shown in the supplementary material of \cite{surace_2020}. In the case of the three states Potts model, we have actually checked that thermalisation and generalised thermalisation give rise to different entanglement spectra also at low energies, at least for the small system sizes we have been able to numerically investigate.  A larger system size analysis is required to make definite statements and for this reason we don't report our results here.

Assuming that in this case, thermalisation and generalised thermalisation result in different entanglement spectra at low entanglement energy, we face an even more complex scenario. The three sate Potts model critical point is not an isolated critical point at zero temperature. As discussed e.g. in \cite{samajdar_2018} the critical landscape, when introducing complex $J$ and $f$ is very involved, and is characterised by a line of transitions \cite{samajdar_2018}. The fact that by quenching the system out of equilibrium we are dealing with a finite-energy-density state, means that we could be noticing a cross-over phenomenon induced by the presence of such a line of critical points. This fact would thus suggest that, even enlarging the system size and going to larger blocks, we would actually never flow to the expected Potts critical point, but we could actually flow towards some different universality class. 

We leave to further studies the task of identifying which of the above scenarios (if any) is responsible for the differences that we have observed between the CFT calculation and the numerical results we have presented for the evolution of the entanglement spectrum  after the quenches to the critical point in the three states Potts model.

\section{Aknowledgments}
We would like to acknowledge the inspiring discussions with E. Tonni and F. Essler on the topics presented.
The numerical computations have been possible thanks to the Tenpy open source package \cite{hauschild_2018}.
L.T. acknowledges support from the Ram\'on y Cajal program RYC-2016-20594, the ``Plan Nacional Generaci\'on de Conocimiento'' PGC2018-095862-B-C22 and and Grant CEX2019-000918-M funded by MCIN/AEI/10.13039/501100011033. This project was supported by the European Union Regional Development Fund within the ERDF Operational Program of Catalunya, Spain (project QUASICAT/QuantumCat, ref. 001- P-001644).

\end{document}